\documentclass[
 reprint,
 amsmath,
 amssymb,
 superscriptaddress,
 aps,
]{revtex4-1}

\usepackage{graphicx}
\usepackage{bm}
\usepackage{dcolumn}
\usepackage{gensymb}
\usepackage[utf8]{inputenc}
\usepackage{float}

\usepackage{array,mathtools,amssymb,booktabs}
\newcolumntype{C}{>{$}c<{$}}
\AtBeginDocument{
\heavyrulewidth=.08em
\lightrulewidth=.05em
\cmidrulewidth=.03em
\belowrulesep=.65ex
\belowbottomsep=0pt
\aboverulesep=.4ex
\abovetopsep=0pt
\cmidrulesep=\doublerulesep
\cmidrulekern=.5em
\defaultaddspace=.5em
}

\usepackage{hyperref}
\newcommand{\tref}[1]{Table~\ref{#1}}

\begin{document}

\title{Role of triple excitations in calculating different properties of Ba$^+$}

\author{S. G. Porsev}
\affiliation{Department of Physics and Astronomy, University of Delaware, Newark, Delaware 19716, USA}
\affiliation{Petersburg Nuclear Physics Institute of NRC ``Kurchatov Institute'', Gatchina, Leningrad District 188300, Russia}
\author{M. S. Safronova}
\affiliation{Department of Physics and Astronomy, University of Delaware, Newark, Delaware 19716, USA}

\begin{abstract}
We carried out calculations of the energies, hyperfine structure constants and electric-dipole transiton amplitudes for the low-lying states
of Ba$^+$ in the framework of the relativistic linearized coupled-cluster single double and coupled-cluster single double (valence)
triple methods. Taking into account that an iterative inclusion of the valence triples into consideration is a complicated and
computationally demanding process we study the effects of computational restriction on the final results. We also present a detailed study of
various corrections to all calculated properties, and we use our results to  formulate several broad  rules that can be used in future calculations
of the elements where experimental data are scarce and correct theoretical predictions are highly important.
\end{abstract}

\date{\today}

\maketitle

\section{Introduction}
Precise values of energies, transition properties, and hyperfine structure (HFS) constants provide fundamental knowledge for atomic and molecular systems and are needed for many applications, including studies of fundamental symmetries and searches for physics beyond the standard model~\cite{SafBudDem18}, development of atomic clocks~\cite{KozSafCre18,LudBoyYe15}, the study of degenerate quantum gases and quantum
information~\cite{CooCovMad18,HeiParSan20,ZhaBisBro14}, plasma physics \cite{MurSafSaf20}, and nuclear physics \cite{PorSafSaf18}.
Theoretical calculations of the hyperfine constants and their evaluated uncertainties are needed to extract nuclear magnetic moments
for nuclear physics studies, especially as rare isotopes will become available with a high yield at the Facility for Rare
Isotope Beams (FRIB) \cite{AbeAviAyr19}.

Comparison with high-precision measurements has a vital role in the development of high-precision theory, as they provide important benchmarks
used to assess theoretical uncertainties.
The theory-experimental comparison carried out here  provides important information for predicting the properties of very heavy elements with
$Z \gtrsim 90$,  where a precision theory is needed to predict energies and matrix elements (MEs) prior to difficult one-atom-at-a-time spectroscopy studies~\cite{PorSafSaf18}.  This work was also motivated by a need to better understand the Th$^{3+}$ properties for the development
of the nuclear clock (see Ref.~\cite{PeiSchSaf21}).

Different properties of Ba$^+$, such as energies, HFS constants, electric-dipole, and parity-nonconserving transition amplitudes,
were repeatedly calculated earlier using various methods
~\cite{EliKalIsh96,DzuFlaGin01,Sah06,Ita06,SahDasCha07,ManAng10,DutMaj14,SafSafRad13,GinVolFri17,ArnChaKae19,ZhaArnCha20,KauDarSah21}.
Since Ba$^+$ is a univalent ion, most of the calculations were carried out in the framework of the relativistic coupled-cluster method, which is
particulary well suited for such cases with only core-core and core-valence correlations.

In the simplest version of this approach only linearized terms involving single ($S$) and double ($D$) excitations
[the linearized coupled-cluster single double (LCCSD) method] are taken into
consideration (for more details, see, e.g., Ref.~\cite{BluJohSap91}).
In the LCCSD method, the initial lowest-order (Dirac-Hartree Fock) wave function
is modified to include all possible single and double excitations from the core and valence electrons to all possible states in the basis set
weighted by the excitation coefficients.
Substituting this modified wave function into the many-body Schr\"odinger equation, one derives a set of iterative equations for the single and double excitation coefficients. Solving these equations gives the correlation energy and a wave function that can be used to compute various MEs.

As a next step, the nonlinear (NL) terms~\cite{EliKalIsh96,RupSafJoh07} or the triple excitations
in the leading order (perturbative triple excitations)~\cite{SafSafRad13,ArnChaKae19} or both of these effects~\cite{Sah06,SahDasCha07} can be
included. The NL single double terms are combinations of the single and double excitations that take the form of $S^2$, $SD$, $D^2$, and
higher-order ($S^3$, etc.) combinations.
Triple valence terms involve excitations of two core electrons and a valence electron, while triple core excitations involve exciting
three core electrons. To the best of our knowledge, there are no calculations for atomic systems where the equations for both the core and valence
triple excitations would be solved iteratively.
The equations for the {\it valence} triples were solved iteratively in very few works~\cite{PorDer06,PorBelDer09,PorBelDer10}, but even
there they were not included in full. Essential restrictions in such a calculation were applied due to high computational demands. Before the present work, an effect of such restrictions was assumed to be small, but not directly evaluated, which is remedied in this work.

Unfortunately, the simple inclusion of more and more correlation corrections
does not guarantee an improved accuracy of the results due to significant cancellations of these terms in certain cases. This cancellation partially explains the exceptional accuracy of the LCCSD approach and its scaled variants for a wide range of properties and systems \cite{SafJoh08}. However, predicting the properties of transition matrix elements and HFS constants with an accuracy below 0.5\%, which becomes essential for many applications, requires inclusions of terms beyond LCCSD. The prior lack of a good set of accurate experimental benchmarks, involving $s$, $p$, and $d$ states,
complicated the development of a high-precision theory.

A next step, an iterative solution of the equations for the core (in addition to the valence) triple excitations, is such a challenging and computationally demanding problem for a heavy system with a large core that it is not clear if it can  be fully implemented even using modern
computer facilities.
Therefore, it is important to establish in what cases we can improve the calculation accuracy including the NL terms and valence triples but disregarding
the core triples and when the core triples cannot be omitted without a loss of accuracy. The LCCSD approach is also used to generate an effective Hamiltonian for systems with several (two to six) valence electrons \cite{SafKozJoh09}. It is important to know for further work with such systems
if extensions of the LCCSD method can reliably improve the calculation accuracy.

A goal of this work is to establish the role of triple excitations in calculating the energies, HFS constants, and $E1$ transition amplitudes
for the low-lying states which can be extended to other systems.
To this end, we carried out calculations in the framework of the coupled-cluster single double (valence) triple (CCSDvT) method. In addition to
the LCCSD and NL terms, the valence triple excitation coefficients are included by solving the respective equations iteratively.
This essentially provides a complete addition of the valence triples with minimal restrictions.  We also studied the effect of omitting inner core shells excitations and limiting the number of partial waves and basis-set states used in the calculations.
We specifically select Ba$^+$ due to the presence of the low-lying metastable $5d_{3/2,5/2}$ states and a
number of recently obtained experimental data~\cite{ArnChaKae19,ZhaArnCha20,ChaKohArn20}.
The available high-precision measurements provide valuable benchmarks of the calculation accuracy that can be achieved within the framework
of such an approach.
\section{Method of calculation}
\label{method}
We evaluated the energies, HFS magnetic dipole constants $A$ of the lowest-lying states, and $E1$ transition amplitudes between the low-lying
levels using a version of the high-precision relativistic coupled-cluster method developed in Ref.~\cite{PorDer06}.
As noted above, the valence triple excitations are included by solving the coupled cluster equations iteratively, whereas core triple excitations are computationally too expansive and are omitted.  Quadratic NL terms ($S^2$, $SD$, and $D^2$) are also included, but cubic and higher-order terms are omitted as these are expected to be smaller than the core triple terms.

Considering Ba$^+$ as a univalent ion, we construct the basis set in the $V^{N-1}$ approximation, where $N$ is the number of electrons.
The initial self-consistency procedure (including the Breit interaction) was carried out for the core electrons, and then the $6s$, $5d$, and $6p$
orbitals were constructed in the frozen-core potential. The remaining virtual orbitals were
formed using 40 basis set B-spline orbitals of order 7 defined on a nonlinear grid with 500 points.
This basis set included partial waves with the orbital quantum number up to $l= 6$.

The coupled cluster equations were solved in a basis set consisting of single-particle states. In the equations for singles and doubles the sums over excited states were carried out with 35 basis orbitals with orbital quantum number $l \leq 6$.

Due to the high computational demands of iterative solutions of the valence triple equations, it is reasonable to estimate the role of different contributions. These results can be used in future calculations for heavier systems, such as Th$^{3+}$, where a full inclusion of the valence
triples  is not possible due to increased core size (a computational time scales as a square of the core shells) and enhanced effects of reducing
the basis set on computational accuracy.  To this end, we solved the equations for the valence triples using several methods of
varying complexity:\\
\noindent (i) the core electron excitations were allowed from the $[3d-5p]$ core shells, the maximal orbital quantum number, $l^{\rm tr}_{\rm max}$,
of all excited orbitals was equal to 4, and the largest principal quantum number $n^{\rm tr}_{\rm max}$ of the virtual
orbitals where excitations were allowed was 25, \\
(ii) the core electrons excitations were allowed from the $[3d-5p]$ shells, with $l^{\rm tr}_{\rm max} =5$,
and $n^{\rm tr}_{\rm max} = 25$,\\
\noindent  (iii) the excitations of {\it all} core electrons were allowed, with $l^{\rm tr}_{\rm max} =4$,
and $n^{\rm tr}_{\rm max} = 25$, and \\
(iv) the excitations of {\it all} core electrons were allowed, with $l^{\rm tr}_{\rm max} =5$,
and $n^{\rm tr}_{\rm max} = 30$.\\
The final (fourth) calculation was most complicated and time-consuming.

As our estimate shows, the remaining contribution, due to the core electrons excitations to $n > 30$ and $l=6$, is small and can be neglected
within the framework of our computational accuracy. As such, the valence triple excitations are practically included 
in full.
\section{Results and discussion}
\subsection{Energies}
Numerical results for the energies are presented in Table~\ref{Tab:E}. The lowest-order Dirac-Hartree-Fock contribution to the energies
is labeled ``BDHF''  since we included the Breit interaction.
$\Delta E_\mathrm{SD}$ is determined as the difference between the LCCSD and BDHF values given in the third and first rows.

As a next step, we carried out calculation in the CCSDvT approximation, including the NL terms and valence triples excitations to get the corrections
$\Delta E_\mathrm{NL}$ and $\Delta E_\mathrm{vT}$, respectively.
We note that NL  terms are included into all triple calculations, i.e., these terms are iterated together. Separate calculation is done with only NL terms, to calculate $\Delta E_\mathrm{NL}$ correction.
 As expected, the correlation corrections ($\Delta E_\mathrm{SD}$,
$\Delta E_\mathrm{NL}$, $\Delta E_\mathrm{vT}$) are dominated by the SD contribution, but both the NL and triple corrections are substantial and do not cancel in this case.
The value $E_{\rm CCSDvT}$ was obtained as $E_{\rm CCSDvT} = E_{\rm LCCSD} + \Delta E_{\rm NL} + \Delta E_{\rm vT}$.

We also found complementary corrections due to the basis extrapolation ($\Delta E_{\rm extrap}$) and
quantum-electrodynamic (QED) radiative corrections ($\Delta E_{\rm QED}$). We determine $\Delta E_{\rm extrap}$ as  the contribution of
the higher ($l > 6$) partial waves. Based on an empiric rule obtained for Ag-like ions in Ref.~\cite{SafDzuFla14PRA1}, we estimate this contribution
as the difference of two calculations carried out with $l_{\rm max} = 6$ and $l_{\rm max} = 5$.

We find that the QED corrections are small; even for the $6s$ state, $\Delta E_{\rm QED}$ is six times smaller in absolute value
than $\Delta E_\mathrm{vT}$. The QED corrections for the $6s$, $5d_{3/2}$, and $5d_{5/2}$ states, calculated in Ref.~\cite{RobDzuFla13},
are in good agreement with our values. The basis extrapolation corrections
are several times smaller in absolute value than $\Delta E_\mathrm{vT}$ for the $6s$ and $6p$ states, but for the $5d_{3/2}$ and $5d_{5/2}$ states
$\Delta E_{\rm extrap}$ are 1.5--2 times larger in absolute value than $\Delta E_\mathrm{vT}$. Thus, an inclusion of the basis extrapolation
corrections is essential for the $5d_{3/2,5/2}$ states. Since $\Delta E_{\rm extrap}$ and $\Delta E_\mathrm{vT}$ are of opposite sign for all
considered states, they tend to cancel each other.

The total theoretical results, designated as $E_{\rm total}$, are obtained as the LCCSD values plus $\Delta E_\mathrm{NL}$, $\Delta E_\mathrm{vT}$,
$\Delta E_{\rm QED}$, and $\Delta E_{\rm extrap}$ corrections.
The experimental energies, taken from the NIST database Ref.~\cite{RalKraRea11}, are given in the row labeled ``$E_{\rm expt}$''.

For calculation of different properties of univalent systems, the LCCSD approximation is used very often and its accuracy is known to be quite satisfactory. As seen from the table, the accuracy of the LCCSD values
is at the level 0.5-0.7\%. When we included into consideration the NL and vT terms, the accuracy of calculation of the energy levels
improved by an order of magnitude. An  agreement between the $E_\mathrm{CCSDvT}$  and experimental values is at the
level of 0.09\% for $6s$ and $6p_{1/2,3/2}$, 0.04\% for $5d_{3/2}$, and 0.02\% for $5d_{5/2}$ (experimental uncertainties are negligible
for comparisons at this level). The addition of the QED and extrapolation corrections results slightly worsens the agreement between
the theoretical ``total'' and experimental values. These final
differences between the experimental and total values are presented in the rows labeled ``Diff. (cm$^{-1}$)''
and ``Diff. (\%)'' in cm$^{-1}$ and \%, respectively. The potential reason for the residual discrepancy is the omitted core triples corrections
that can partially cancel the complementary QED and extrapolation corrections.
\begin{table}[h]
\caption{Contributions to removal energies of the $6s$, $5d_{3/2}$, $5d_{5/2}$, $6p_{1/2}$, and $6p_{3/2}$ states for Ba$^+$ (in cm$^{-1}$) in
different approximations, discussed in the text, are presented. The theoretical CCSDvT values and experimental results
are given in the rows $E_{\rm CCSDvT}$  and $E_{\rm expt}$. The differences between the experimental and CCSDvT values are presented
in the rows labeled ``Diff. (cm$^{-1}$)'' and ``Diff. (\%)'' in cm$^{-1}$ and \%, respectively.
The results of other works are listed in the bottom panel for comparison.}
\label{Tab:E}
\begin{ruledtabular}
\begin{tabular}{lrrrrr}
\smallskip
                                        &   $6s$   & $5d_{3/2}$ &    $5d_{5/2}$    & $6p_{1/2}$ & $6p_{3/2}$ \\
\hline \\[-0.7pc]
$E_{\rm BDHF}$                          &  75326   &   68183    &      67727       &   57237    &   55861  \\[0.3pc]
$\Delta E_{\rm SD}$                     &   5916   &    8173    &       7804       &    3488    &    3148  \\[0.1pc]
$E_{\rm LCCSD}$                         &  81242   &   76356    &      75532       &   60725    &   59009 \\[0.4pc]

$\Delta E_{\rm NL}$                     & $-$326   &  $-$474    &     $-$458       &  $-$239    &  $-$222  \\[0.1pc]
$\Delta E_{\rm vT}$                     & $-$291   &  $-$103    &      $-$77       &  $-$115    &  $-$103  \\[0.1pc]

$E_{\rm CCSDvT}$                        &  80625   &   75779    &      74997       &   60371    &   58684 \\[0.4pc]
\multicolumn{5}{l}{\it Complementary corrections:} \\
{$\Delta E_{\rm QED}$}$^{\rm a}$        &  $-$46   &      20    &         19       &       4    &        4 \\[0.1pc]
$\Delta E_{\rm extrap}$                 &     26   &     163    &        157       &      20    &       19 \\[0.2pc]
\hline \\[-0.6pc]
$E_{\rm total}$                         &  80605   &   75962    &      75173       &   60394    &   58707  \\[0.3pc]
$E_{\rm expt}$~\cite{RalKraRea11}       &  80686   &   75812    &      75011       &   60425    &   58734  \\[0.1pc]
Diff. (cm$^{-1}$)                       &     82   &  $-$150    &     $-$162       &      31    &      27  \\[0.1pc]
Diff. (\%)                              &    0.1   &    -0.2    &       -0.2       &    0.05    &    0.05  \\
Other works:                            &&&&& \\
Eliav {\it et al.}~\cite{EliKalIsh96}   &  80871   &   75603    &      74778       &   60475    &   58768  \\[0.1pc]
Dzuba {\it et al.}~\cite{DzuFlaGin01}   &  80813   &   76402    &      75525       &   60581    &   58860  \\[0.1pc]
Sahoo {\it et al.}~\cite{SahDasCha07}   &  80794   &   75481    &      74346       &   60384    &   58690  \\[0.1pc]
Roberts {\it et al.}~\cite{RobDzuFla13a}&  80838   &   76558    &      75710       &   60604    &   58878
\end{tabular}
\end{ruledtabular}
\begin{flushleft}
$^{\rm a}${The QED corrections for the $6s$, $5d_{3/2}$, and $5d_{5/2}$ states, calculated in Ref.~\cite{RobDzuFla13}, are in a good agreement with our values}.
\end{flushleft}
\end{table}

In Sec.~\ref{method} we described four variants of an inclusion of the valence triples. In \tref{E:trip} we give
the valence triple contributions to the energies obtained in different approximations. In the first row of the table we present the valence triple
correction, designated as $\Delta E_{\rm vT1}$, obtained when only the core electrons excitations from the $[3d-5p]$ shells were allowed,
$l^{\rm tr}_{\rm max} =4$, and $n^{\rm tr}_{\rm max} = 25$. In rows 2-4 we separately give the additional corrections due to:
(1) increasing $l^{\rm tr}_{\rm max}$ from 4 to 5, (2) the inclusion of the excitations from the $[1s-3p]$ core shells, and
(3) the inclusion of the excitations to the virtual orbitals with $n= 25-30$. The sum of values in rows 1-4 give the total correction
designated as ``$\Delta E_{\rm vT}$''. We find that  the additional  inclusion of the $l=5$ partial wave is significant for the $5d$ states,
and the initial $\Delta E_{\rm vT1}$ approximation overestimates the contribution of the valence triples to the $5d_{5/2}$ energy by a factor of 2.
\begin{table}[h]
\caption{The triple (vT) contributions to the energies in cm$^{-1}$. The vT
correction, labeled as ``$\Delta E_{\rm vT1}$'' and obtained when the core electrons excitations are allowed from the $[3d-5p]$ shells,
$l^{\rm tr}_{\rm max} =4$, and $n^{\rm tr}_{\rm max} = 25$, is given in the first row. In the rows 2-4, we separately give the additional corrections
due to: (1) increasing $l^{\rm tr}_{\rm max}$ from 4 to 5, (2) the inclusion of the excitations from the $[1s-3p]$ core shells, and
(3) the inclusion of the excitations to the virtual orbitals with $n= 25-30$. The sum of the values in rows 1-4 gives the total vT correction,
$\Delta E_{\rm vT}$.}
\label{E:trip}
\begin{ruledtabular}
\begin{tabular}{lrrrrr}
\smallskip
                            &   $6s$   & $5d_{3/2}$ &    $5d_{5/2}$    & $6p_{1/2}$ & $6p_{3/2}$ \\
\hline \\[-0.7pc]
$\Delta E_{\rm vT1}$        &  -315    &    -185    &       -161       &    -136    &    -123    \\[0.3pc]
$l^{\rm tr} = 5$            &    14    &      65    &         66       &      14    &      14    \\[0.1pc]
$[1s-3p]$ excs.             &    13    &      30    &         29       &       9    &       8    \\[0.4pc]
$n^{\rm tr} = 26-30$        &    -3    &     -12    &        -13       &      -3    &      -2    \\[0.4pc]
 $\Delta E_{\rm vT}$        &   -291   &    -103    &        -78       &    -115    &    -103    \\[0.1pc]
\end{tabular}
\end{ruledtabular}
\end{table}

We find that numerically the vT corrections are smaller than the NL corrections. What is more important is that
these corrections are of the same sign for all five states considered here. Based on a comparison between the experimental and
theoretical results, we conclude that in such a case the inclusion of the NL and vT terms is completely justified and significantly improves
the agreement between theory and experiment. Note that this is similar to what was observed in Cs in Ref.~\cite{PorBelDer10},
where the consideration was limited by the low-lying $S$ and $P_{1/2}$ states.
\subsection{Magnetic dipole HFS constants}
The magnetic dipole HFS constants $A$ were calculated for 137 isotope of Ba$^+$ with the nuclear spin $I=3/2$ and the nuclear magnetic moment
$\mu = 0.93737(2)\, \mu_N$ (where $\mu_N$ is the nuclear magneton)~\cite{Sto05}.
We chose this isotope for a benchmark calculation because its nuclear magnetic moment is known with high accuracy and there are high-precision
experimental data for the HFS constants $A$.

The expression for the  magnetic dipole HFS constant $A$ of an atomic state $\left| J\right\rangle $ can be written as
\begin{eqnarray}
A = \frac{g_N}{\sqrt{J(J+1)(2J+1)}} \langle  J ||T || J \rangle .
\end{eqnarray}
Here $g_N = \mu/(\mu_N\,I)$ and $\langle  J || T || J \rangle$ is the reduced matrix element of
the operator ${\bf T}$. Assuming the nucleus to be a charged ball of uniform magnetization with radius $R$, the single-particle operator
${\bf T}$ can be written as (we use atomic units $\hbar=|e|=m=1, \, c \approx 137$)
\begin{eqnarray*}
{\bf T} = \frac{ {\bf r} \times \boldsymbol{\alpha}}{c\,r_{>}^3}\, \mu_N,
\end{eqnarray*}
where $\boldsymbol{\alpha}$ is the Dirac matrix, $r$ is the radial position of the valence electron,
${\bf r} \times \boldsymbol{\alpha}$ is the vector product of ${\bf r}$ and $\boldsymbol{\alpha}$,
and
\begin{eqnarray}
 r_> \equiv
\left\{
\begin{array}{r}
r,\,\,{\rm if}\,\, r \geq R,\\
R,\,\,{\rm if}\,\, R > r .
\end{array}
\right.
\end{eqnarray}

The results for the HFS constants are presented in \tref{Tab:Ahfs}.
\begin{table}[h]
\caption{Different contributions to the magnetic-dipole hyperfine-structure constants $A$ (in MHz) for $^{137}$Ba$^+$.
Results of calculations and comparison with the experimental values are presented. See the text for the explanation of entries.}
\label{Tab:Ahfs}
\begin{ruledtabular}
\begin{tabular}{lrrrrr}
                         & $A(6s)$          & $A(5d_{3/2})$   & $A(5d_{5/2})$    & $A(6p_{1/2})$ & $A(6p_{3/2})$ \\
\hline \\[-0.6pc]
BDHF                     &   2912           & 128.4           &   51.6           &  491.0           &    71.8        \\[0.4pc]

$\Delta$(SD)             &   1274           &  68.3           &  -61.1           &  276.4           &    55.5        \\[0.1pc]
LCCSD                    &   4186           & 196.7           &   -9.5           &  767.3           &   127.3        \\[0.4pc]

$\Delta$(NL)             &    -67           &  -5.3           &    2.8           &  -26.2           &    -3.3        \\[0.1pc]
$\Delta$(vT)             &    -34           &  -3.1           &   -8.4           &    3.4           &     2.3        \\[0.1pc]

CCSDvT                   &   4085           & 188.4           &  -15.1           &  744.6           &   126.3        \\[0.3pc]
\multicolumn{5}{l}{\it Complementary corrections:} \\
$\Delta$(scale)          &     10           &   0.1           &    0.02          &    3.0           &     0.5        \\[0.1pc]
Ln \& vtx dres.          &     -7           &  -1.3           &    0.9           &   -2.8           &     0.0        \\[0.1pc]
QED                      &    -15           &                 &                  &                  &                \\[0.1pc]
Basis extrap. 1          &      4           &   0.6           &    0.5           &    0.6           &     0.1        \\[0.1pc]
Basis extrap. 2          &    -23           &  -0.8           &    0.8           &   -1.6           &    -0.7        \\[0.2pc]
\hline \\[-0.6pc]
Total                    &   4054           & 186.9           &  -12.9           &  743.8           &   126.1        \\[0.2pc]
Experiment               &   4019$^{\rm a}$ & 189.7$^{\rm b}$ &  -12.0$^{\rm c}$ &  743.7$^{\rm d}$ &   127.2$^{\rm e}$  \\[0.2pc]
Difference               &  -0.9\%          &  1.5\%          &  -7.7\%          & -0.02\%          &   0.7\%        \\[0.4pc]

Other works:             &&&&& \\
Ref.~\cite{SafSafRad13}  &   3998           &    191.5        &  -10.0           &  734.0           &   121.3        \\[0.2pc]
Ref.~\cite{DutMaj14}     &   4112           &    194.2        &                  &  731.1           &   123.1
\end{tabular}
\end{ruledtabular}
\begin{flushleft}
$^{\rm a}$Reference~\cite{BlaWer81}, $A= 4018.8708343(16)\, {\rm MHz}$; \\
$^{\rm b}$Reference~\cite{LewChuCaz13}, $A= 189.731494(17)\, {\rm MHz}$; \\
$^{\rm c}$Reference~\cite{LewChuCaz13}, $A= -12.02934(11)\, {\rm MHz}$; \\
$^{\rm d}$Reference~\cite{VilArnHei93}, $A= 743.7(0.3)\, {\rm MHz}$; \\
$^{\rm e}$Reference~\cite{VilArnHei93}, $A= 127.2(0.2)\, {\rm MHz}$.
\end{flushleft}

\end{table}
The LCCSD and BDHF values and the difference between them,  $\Delta$(SD), are given in the upper panel of the table.
The rows 4 and 5 give the corrections due to the NL terms, $\Delta$(NL), and the valence triples, $\Delta$(vT).
The CCSDvT values, obtained as the sum of the LCCSD values and the NL and vT corrections, are presented in the row
labeled ``CCSDvT''. A breakdown of the vT contributions, similar to that presented for the energies, is given for the HFS constants
in \tref{A:trip}.
\begin{table}[h]
\caption{The vT contributions to the HFS constants in MHz. The vT
correction, labeled as ``$\Delta({\rm vT1})$'' and obtained when the core electrons excitations are allowed from the $[3d-5p]$ shells,
$l^{\rm tr}_{\rm max} =4$, and $n^{\rm tr}_{\rm max} = 25$, is given in the first row. In rows 2-4, we give the additional corrections
due to: (1) increasing $l^{\rm tr}_{\rm max}$ from 4 to 5, (2) the inclusion of the excitations from the $[1s-3p]$ core shells, and
(3) the inclusion of the excitations to the virtual orbitals with $n= 25-30$. The sum of the values in rows 1-4 gives the total vT correction,
$\Delta({\rm vT})$.}
\label{A:trip}
\begin{ruledtabular}
\begin{tabular}{lrrrrr}
\smallskip
                            & $A(6s)$  & $A(5d_{3/2})$ &  $A(5d_{5/2})$   & $A(6p_{1/2})$ & $A(6p_{3/2})$ \\
\hline \\[-0.7pc]
$\Delta({\rm vT1})$         &   -43    &      -3.9     &        -8.0      &      0.9      &     1.7    \\[0.3pc]
$l^{\rm tr} = 5$            &     3    &       0.5     &        -0.01     &      1.0      &     0.3    \\[0.1pc]
$[1s-2p]$ excs.             &     3    &       0.3     &        -0.4      &      1.7      &     0.4    \\[0.4pc]
$n^{\rm tr} = 26-30$        &     3    &      -0.1     &         0.01     &     -0.2      &   -0.02    \\[0.4pc]
 $\Delta({\rm vT})$         &   -34    &      -3.1     &        -8.4      &      3.4      &     2.3    \\[0.1pc]
\end{tabular}
\end{ruledtabular}
\end{table}

There are also complimentary corrections. Since the CCSDvT method is an approximation and we miss certain contributions
(for instance, the contribution of the core triples), it leads to a difference between the computed and experimental energies.
To partially account for the missing contributions in calculations of matrix elements, we additionally
correct the valence singles obtained at the CCSDvT stage, rescaling them by the ratio of experimental and theoretical correlation energies.
Such a semiempirical procedure was suggested in Ref.~\cite{BlaSapJoh92}, and in the following we will refer to results
obtained using it as ``scaling''. In addition to the correction arising due to scaling, $\Delta$(scale),
we calculated the line and vertex dressing corrections discussed in detail in~\cite{DerPor05} and labeled in \tref{Tab:Ahfs} as ``Ln \& vtx dres.''.

The QED corrections and the corrections due to the basis-set extrapolation are also given on the respective lines of the table.
For the $6s$ state, we estimated the QED correction to be -16 MHz.
A rigorous calculation of the one-loop QED radiative corrections to the ground state hyperfine structure interval for $^{135}$Ba$^+$ was
carried out in Ref.~\cite{GinVolFri17}, and its relative contribution was found to be (-0.38\%). Applying this result to our case of $^{137}$Ba$^+$,
we find the QED correction to $A(6s)$ to be -15 MHz, in agreement with our result. According to our estimate, the QED corrections for all
other HFS constants are two orders of magnitude smaller and we neglect them.

The basis extrapolation correction consists of two parts in this case, labeled in the table as ``Basis extrap. 1'' and
``Basis extrap. 2''. To find them, we additionally constructed a longer basis set, consisting of 50 B-spline orbitals and, as previously,
including partial waves with the orbital quantum number up to $l = 6$.
Using this basis set, we calculated the HFS constants in the LCCSD approximation with $l_{\rm max} = 6$ and $l_{\rm max} = 5$.
The first correction, ``Basis extrap. 1'', accounts for the contribution of the
partial waves with $l > 6$, and we estimated it as the difference between the values obtained for $l_{\rm max} = 6$ and $l_{\rm max} = 5$.
The ``Basis extrap. 2'' correction is estimated as the difference between the LCCSD values obtained for 50 and 40 B-spline basis sets
at $l_{\rm max} = 6$. We note that the second basis extrapolation correction is essential only for the HFS constants and gives a negligible
contribution to the energies and the electric dipole MEs, which we will discuss below.
A full calculation of the valence triple corrections using the 50 B-spline basis set is presently untractable, and the relative correction
at the LCCSD level is small. For this reason, a shorter (40 B-spline) basis set was used.

The total values (labeled as ``Total'') are obtained as the sum of the CCSDvT values and all complementary corrections.
The difference between the experimental and ``Total'' values is given in the row labeled ``Difference''.
The results of other works are presented in the lower panel of the table for comparison.

Comparing the LCCSD values with the experiment, we see that the agreement for the $A(6s)$, $A(5d_{3/2})$, and $A(6p_{1/2})$ is at the level of 3-4\%.
The HFS constant $A$ for $5d_{5/2}$ differs from the experiment by 21\%. This is not surprising because the $\Delta$(SD)
correction is huge and even changes the sign of the BDHF value. Based on this, we can expect a high sensitivity of  $A(5d_{3/2})$
to other corrections as well. An almost perfect agreement of the LCCSD and experimental values of $A(6p_{3/2})$ appears to be fortuitous.

At the CCSDvT stage (after including the NL and vT corrections), the agreement between the theoretical and experimental results
for $A(6s)$, $A(5d_{3/2})$, and $A(6p_{1/2})$ noticeably improved to the level of 0.1-1.6\%.
Such an improvement is similar to what we observed for the energies. Further inclusion of the complementary corrections changed
all the HFS constants except $A(6s)$ only slightly. The HFS constant $A(6s)$ is more sensitive to the QED and basis extrapolation corrections
than the other HFS constants. It changed by 1\% and its total value is in 0.9\% agreement with the experiment.
We note also that we consider the nucleus as the uniformly magnetized ball. As was shown in the recent
paper~\cite{RobGin21}, using a more sophisticated modeling of the magnetization distribution that takes into account the nuclear angular momenta and
configuration may lead to an additional correction to $A(6s)$ on the order of 0.3\%.

Analyzing the results, we see that the NL and vT contributions for $A(6s)$ and $A(5d_{3/2})$ are of the
same sign, and their CCSDvT values are closer to the experiment than the LCCSD results.
For $A(6p_{1/2})$ the vT correction is relatively small; it is an order of magnitude smaller in absolute value than the NL correction, and it does
not play an essential role. For $A(6p_{3/2})$ a very good agreement with the experimental
value was achieved already at the LCCSD stage. So, we can expect that all other corrections should essentially cancel each other.
This is confirmed by the results presented in the last column of \tref{Tab:Ahfs}.
We assume that a deviation of the total value of $A(6p_{3/2})$ from the experimental result is due to
omitted core triple correction. Finally, $A(5d_{5/2})$ is small and sensitive to different corrections.
For these reasons, its accurate calculation is difficult. The vT correction is relatively large, and therefore one can expect a non-negligible
contribution from the core triples. Taking all this into account, we consider the agreement of the total value with the experiment at the level
of 8\% as satisfactory.
\subsection{Electric dipole matrix elements}
The $E1$ matrix elements were calculated previously for transitions between the low-lying states in Ref.~\cite{ArnChaKae19}
in the framework of different approximations including scaling and perturbative triples. Combining them with the experimental measurements,
the authors of Ref.~\cite{ArnChaKae19} provided the recommended values for a number of the reduced matrix elements of the electric dipole operator
${\bf D} = - {\bf r}$ for the transitions between the low-lying states.

Applying the CCSDvT approach, we calculated a number of the MEs of the $E1$ operator.
The results are presented in \tref{Tab:E1}.
The values listed in the row labeled ``LCCSD'' are obtained in the LCCSD approximation, and
the SD correction, $\Delta$(SD), is the difference between the LCCSD and BDHF values. On the next line, labeled
``Difference (LCCSD-Recomm.)'', the difference between the LCCSD and recommended values from Ref.~\cite{ArnChaKae19} is given in \%.
We note a very good agreement (0.2-0.4\%) with the experiment at the LCCSD stage. As we will show below, in contrast with the energies and
HFS constants, an inclusion of other corrections does not improve this agreement.

Rows 4-6 give the NL and vT corrections and the CCSDvT value is obtained as LCCSD + $\Delta$(NL) + $\Delta$(vT).
The vT correction in this case is completely dominated by $\Delta$(vT1), obtained when the core electrons excitations are allowed from the $[3d-5p]$ shells, $l^{\rm tr}_{\rm max} =4$, and $n^{\rm tr}_{\rm max} = 25$; it contributes more than 85\% to $\Delta$(vT). For this reason, we do not
present the breakdown of the vT corrections for the $E1$ transition amplitudes.

We see that in this case the NL and vT corrections are of opposite sign for all considered MEs, and $\Delta$(vT)
are 2-2.5 times larger in absolute value than $\Delta$(NL). Since the contribution of the vT corrections is relatively large,
we can expect that a contribution of the core triples (omitted here and in all other works) can noticeably change the total results.

The complementary corrections, including the scaling, line, and vertex dressing, QED, and basis extrapolation corrections, are listed
in the respective rows of \tref{Tab:E1}. The QED corrections for $|\langle 6p_{1/2} ||D|| 6s \rangle|$ and $|\langle 6p_{3/2} ||D|| 6s \rangle|$ are
very small. We found them to be 0.002 and 0.003 a.u., respectively. These values are in very good agreement with those obtained in
Ref.~\cite{RobDzuFla13}. According to our estimate, the QED corrections for other matrix elements listed in \tref{Tab:E1} are even smaller, and
we neglect them.
Since the complementary corrections are small and, in addition, they partially cancel each other, their total
contribution changes the matrix elements only slightly. The total results and their differences with the recommended values are
given on the lines labeled ``Total'' and ``Difference (Total-Recomm.)''. The accuracy of the total values is at the level of 0.8-1.3\%.
Potentially a deterioration of the accuracy of the ``Total'' values compared to the LCCSD values can be attributed to the omitted core triples.
\begin{table*}[htp]
\caption{Different contributions to the reduced matrix elements of the $E1$ operator (in $|e|\,a_B$, where $a_B$ is the Bohr radius). Absolute values are given.
Results of calculations and comparison with the recommended values are presented. See the text for the explanation of entries.}
\label{Tab:E1}
\begin{ruledtabular}
\begin{tabular}{lccccc}
                              & $|\langle 6p_{1/2} ||D|| 6s \rangle|$
                                                  & $|\langle 6p_{3/2} ||D|| 6s \rangle|$
                                                                   & $|\langle 6p_{1/2} ||D|| 5d_{3/2} \rangle|$
                                                                                        & $|\langle 6p_{3/2} ||D|| 5d_{3/2} \rangle|$
                                                                                                           & $|\langle 6p_{3/2} ||D|| 5d_{5/2} \rangle|$\\
\hline \\[-0.6pc]
BDHF                           &   3.892          &  5.479           &  3.741           &  1.634           &    4.993       \\[0.4pc]

$\Delta$(SD)                   &  -0.554          & -0.769           & -0.690           & -0.301           &   -0.890       \\[0.1pc]
LCCSD                          &   3.338          &  4.709           &  3.050           &  1.332           &    4.103       \\[0.4pc]
Difference (LCCSD-Recomm.)     &   0.4\%          &   0.2\%          &  0.3\%           &  0.3\%           &    0.3\%        \\[0.4pc]

$\Delta$(NL)                   &   0.028          &  0.039           &  0.035           &  0.015           &    0.049       \\[0.1pc]
$\Delta$(vT)                   &  -0.068          & -0.091           & -0.075           & -0.031           &   -0.095       \\[0.1pc]
CCSDvT                         &   3.298          &  4.657           &  3.010           &  1.317           &    4.057       \\[0.3pc]

\multicolumn{5}{l}{\it Complementary corrections:} \\
$\Delta$(scale)                &  -0.002          & -0.003           & -0.001           & -0.001           &    0.000       \\[0.1pc]
Ln \& vtx dres.                &   0.002          &  0.002           &  0.002           &  0.001           &    0.002       \\[0.1pc]
QED$^{\rm a}$                  &   0.002          &  0.003           &                  &                  &                \\[0.1pc]
Basis extrap.                  &  -0.001          & -0.001           & -0.009           & -0.004           &   -0.012       \\[0.2pc]
\hline \\[-0.6pc]
Total                          &   3.299          &  4.659           &  3.001           &  1.312           &    4.047       \\[0.2pc]
Recommended~\cite{ArnChaKae19} &  $3.3251(21)$    & $4.7017(27)$     & $3.0413(21)$     & $1.3285(13)$     &   $4.0911(31)$  \\[0.2pc]
Difference (Total-Recomm.)     &   0.8\%          &   0.9\%          &  1.3\%           &  1.2\%           &    1.1\%        \\[0.4pc]

Other works:                   &&&&& \\
Ref.~\cite{KauDarSah21}        &  $3.33(3)$       & $4.71(5)$        & $3.07(2)$        & $1.34(1)$        &   $4.13(3)$    \\[0.2pc]
Ref.~\cite{ZhaArnCha20}        &                  &                  &                  & $1.3320(10)$     &   $4.1028(25)$ \\[0.2pc]
\end{tabular}
\end{ruledtabular}
\begin{flushleft}
$^{\rm a}${The QED corrections for $|\langle 6p_{1/2} ||D|| 6s \rangle|$ and $|\langle 6p_{3/2} ||D|| 6s \rangle|$, calculated
in Ref.~\cite{RobDzuFla13}, are in a good agreement with our values}.
\end{flushleft}
\end{table*}
\section{Conclusion and final remarks}
We carried out calculations of the energies, HFS constants, and $E1$ transition amplitudes for the low-lying states
in the framework of the relativistic CCSDvT method. We have analysed the role of different contributions including the quadratic NL terms
and the iterative valence triples. Taking into account that an iterative inclusion
of the valence triples into consideration is a complicated and computationally demanding process, it is important to be able to predict when the resulting values are expected to be of a higher accuracy than much faster to run the LCCSD and its scaled variants. We note that NL corrections are not as computationally intensive as triples, and restricted smaller triple runs are sufficient to establish if much lengthy full-scale triple calculation is needed to produce high-accuracy recommended values.
Below we formulate several rules
that can be used in future calculations of the elements where experimental data are scarce and correct theoretical predictions are
highly important.

Based on our analysis we conclude the following:
\begin{itemize}
\item
(i) If both the NL and vT contributions are of the same sign and the vT contribution is smaller in absolute value than the NL contribution, an inclusion
of the valence triples is justified because  it is expected to improve the calculation accuracy.
In this case, a sum of the NL and vT corrections is sufficiently large, so that
the LCCSD values can change significantly. For the same reason, the role of omitted core triples is expected to be relatively small.
Their contribution should be smaller than that of the valence triples, and their inclusion
will change the final result only slightly.

\item (ii) If the vT and NL contributions are of the opposite sign and especially if the former
is larger (in absolute value), an inclusion of the vT (and NL) terms into consideration cannot guarantee am improvement of the accuracy
compared to the LCCSD results (as we observed for the $E1$ matrix elements).
In this case, the vT and NL corrections partially cancel each other, and a role of all other omitted corrections increases.
If the vT corrections are larger in absolute value than the NL corrections, it means that the calculated property is rather sensitive to
the triples corrections, and the contribution from the core triples may be substantial.

\item(iii) Based on the analysis of the results obtained, we assume that a future inclusion of the core triples (and potentially other nonlinear terms)
has to be pursued to {\it systematically} improve the predictive accuracy of different observables to a level of better than 0.5\%.
\end{itemize}
\section{Acknowledgements}
We are grateful to Murray Barrett for reading the manuscript and useful remarks.
This work is a part of the ``Thorium Nuclear Clock'' project that  has received funding from the European Research Council  (ERC) under
the European Union's Horizon 2020 research and innovation program (Grant Agreement No. 856415).
S.P. acknowledges support by the Russian Science Foundation under Grant No. 19-12-00157.

%

\end{document}